# Optimal Visual Search with Highly Heuristic Decision Rules

Anqi Zhang and Wilson S. Geisler
University of Texas at Austin

**Abstract**

Visual search is a fundamental natural task for humans and other animals. We investigated the decision processes humans use in covert (single fixation) search with briefly presented displays having well-separated potential target locations. Performance was compared with the Bayesian-optimal decision process under the assumption that the information from the different potential target locations is statistically independent. Surprisingly, humans performed slightly better than optimal, despite humans' substantial loss of sensitivity in the fovea ("foveal neglect"), and the implausibility of the human brain replicating the optimal computations. We show that three factors can quantitatively explain these seemingly paradoxical results. Most importantly, simple and fixed heuristic decision rules reach near optimal search performance. Secondly, foveal neglect primarily affects only the central potential target location. Finally, spatially correlated neural noise can cause search performance to exceed that predicted for independent noise. These findings have broad implications for understanding visual search tasks and other identification tasks in humans and other animals.

**Introduction**

A fundamental and ubiquitous visual task is searching the local environment for specific targets. Humans and other primates have foveated visual systems with high spatial resolution in the direction of gaze (in the fovea) and rapidly declining resolution away from the direction of gaze (in the periphery). Thus, visual search typically involves a series of fixations. During each fixation, the scene is covertly searched to detect the target and/or possible target locations. A sensible and common research strategy is to begin by studying the mechanisms of covert search and then generalize to search with eye movements (overt search). This study focuses on covert (single fixation) search.

Carefully controlled studies of covert search typically present the stimuli briefly to prevent eye movements and with the potential target locations arrayed at a fixed distance from the fixation location to hold the target visibility at the different locations approximately constant (Carrasco, 2011; Eckstein, 2011; Eckstein et al., 2000; Palmer et al., 1993; Palmer & Davis, 2004). In some studies, the task is simply to indicate whether the target is present or absent. In other studies, the target is always present, and the task is to indicate the location of the target. The covert search task in our study is a bit more complex and realistic: identify whether the target is present or absent, and if present, indicate its location.

Under natural search conditions, the number of potential target locations often varies from one situation to the next and hence varying the number of potential locations is a key experimental manipulation. When there is just a single potential target location, the search task reduces to a simple identification, discrimination, or detection task. In general, as the number of potential target locations increases, search accuracy and speed decrease (for reviews see Carrasco, 2011; Eckstein, 2011). The major scientific questions are what stimulus and neural factors are responsible for the decreases and whether models that incorporate the relevant neural factors can quantitatively predict search performance.

Bayesian statistical decision theory provides a principled computational framework for addressing these questions. Specifically, measurements of detection performance for each potential target location, when the location is known, can be used to predict quantitatively the best possible (optimal) performance in the search task when the location of the target is unknown. These predictions provide the normative benchmark for evaluating the various potential stimulus and neural factors affecting search performance. For example, hypothesized factors that cause the Bayesian observer's performance to fall below the measured human performance can be confidently rejected.

As mentioned above, in many well-controlled covert-search studies the targets are presented briefly, and the potential target locations are at fixed retinal eccentricity. Studies show that when the task is to report whether a single target is present or absent, there is a fairly wide range of conditions where human search accuracy is consistent with an optimal decision rule given statistical independence at the potential target locations (Eckstein, 1998, 2011; Eckstein et al., 2000; Palmer et al., 1993). However, fixed eccentricity displays are not representative of natural search, where potential target locations are typically more uniformly distributed across the visual field. Also, a target's visibility typically varies somewhat around a circle at fixed retinal eccentricity (Carrasco et al., 1995; Carrasco & McElree, 2001; Najemnik & Geisler, 2005; Rovamo & Virsu, 1979). Finally, relatively few studies have used backgrounds with dense random noise, which is more typical in natural conditions. Most of the studies that measured search in noise backgrounds allowed multiple fixations (Burgess & Ghandeharian, 1984; Bochud et al., 2004; Najemnik & Geisler, 2005).

Therefore, we designed covert search tasks in white noise backgrounds for target locations that cover the central 16 deg of the visual field. The stimuli were presented for 250 msec—the typical duration of fixations during overt search (Greene, 2006; Hooge & Erkelens, 1998). Search performance was measured in human observers. To increase interpretability, we also directly measured the detectability of the target at each potential target location when the target location was known to the observer (cued). Furthermore, we carefully interleaved the single-cued-location and search sessions to eliminate differences in practice effects for the two kinds of sessions.

The results are surprising. First, all four observers performed the search task slightly better than the Bayesian optimal searcher, given the measured detectability when the target locations were cued and the assumption of statistical independence of the responses at the different locations. Second, the Bayes optimal searcher takes into account the prior probability of the target being present at each potential location (the "prior map"), as well as the detectability of the target at each potential target location (the "$d'$ map"). It seems implausible that during the experiment the observers could precisely learn the prior map and their own $d'$ map, and then optimally apply this information in making responses. Third, in the search task, all four observers showed a substantial loss of sensitivity in the fovea ("foveal neglect"), a phenomenon we reported in a recent study of covert search in continuous noise backgrounds (Walshe & Geisler, 2022).

We show that three factors can explain our seemingly paradoxical results. First, we show that it is not necessary to know accurately the prior and $d'$ maps. Extremely crude and fixed heuristic decision rules, in combination with local normalization (e.g., luminance and contrast gain control), are sufficient to obtain effectively optimal performance. This finding differs from the assumptions in much of the perceptual decision-making literature concerned with understanding the how the brain takes into account the stimulus reliability and prior probabilities (Geisler, 2011; Knill & Richards, 1996; Ma et al., 2022). Second, foveal neglect primarily affects only the central target location. Third, spatially correlated noise corresponding to about 45% of the total noise variance is sufficient to predict the slightly supraoptimal performance, even with the foveal neglect and the heuristic decision rules.

## Methods

Carefully controlled experimental conditions were used to compare performance in covert search with that in simple detection where the potential location of the target is cued. The stimuli in the experiments were generated with MATLAB 2023a and the Psychophysics Toolbox (Brainard, 1997; Pelli, 1997). They had a resolution of 30 pixels per visual degree, with each pixel occupying a 2 x 2 screen pixel region on a calibrated Sony GDM-FW900 cathode-ray-tube (CRT) monitor, with a size of 1920 x 1200 pixels, a refresh rate of 85 Hz, and a bit depth of 8. The stimuli were gamma-compressed prior to display on the screen.

The stimulus presentation sequence for both the search and detection experiments is shown in Figure 1. On each trial, the observer fixated the central circle in a cueing display of faint circles showing the 19 possible target locations. After a brief period of blank screen, the stimulus display was presented for 250 msec, followed by a response period of up to 3000 msec, during which the cueing display was again presented. In the simple detection experiment (Task 1), the potential location of the target was cued, before and after the stimulus presentation, with a low contrast dark circle, with light circles at all other locations. All 19 potential target locations contained at 3.5° diameter randomly sampled patch of high contrast (20% rms contrast) Gaussian white noise. The target, if present, was a small wavelet (6 cycles/deg), windowed to a diameter of 0.8 deg, and added to the center of the noise patch. The luminance outside the circular cues was set to the mean luminance of stimuli within the cues (60 $cd/m^2$). Light circular cues had a luminance of 66 $cd/m^2$. Dark circular cues had a luminance of 51 $cd/m^2$. The search task (Task 2) was identical except the target (if present) could appear any location with equal probability, and all potential target locations were indicated with light circles.

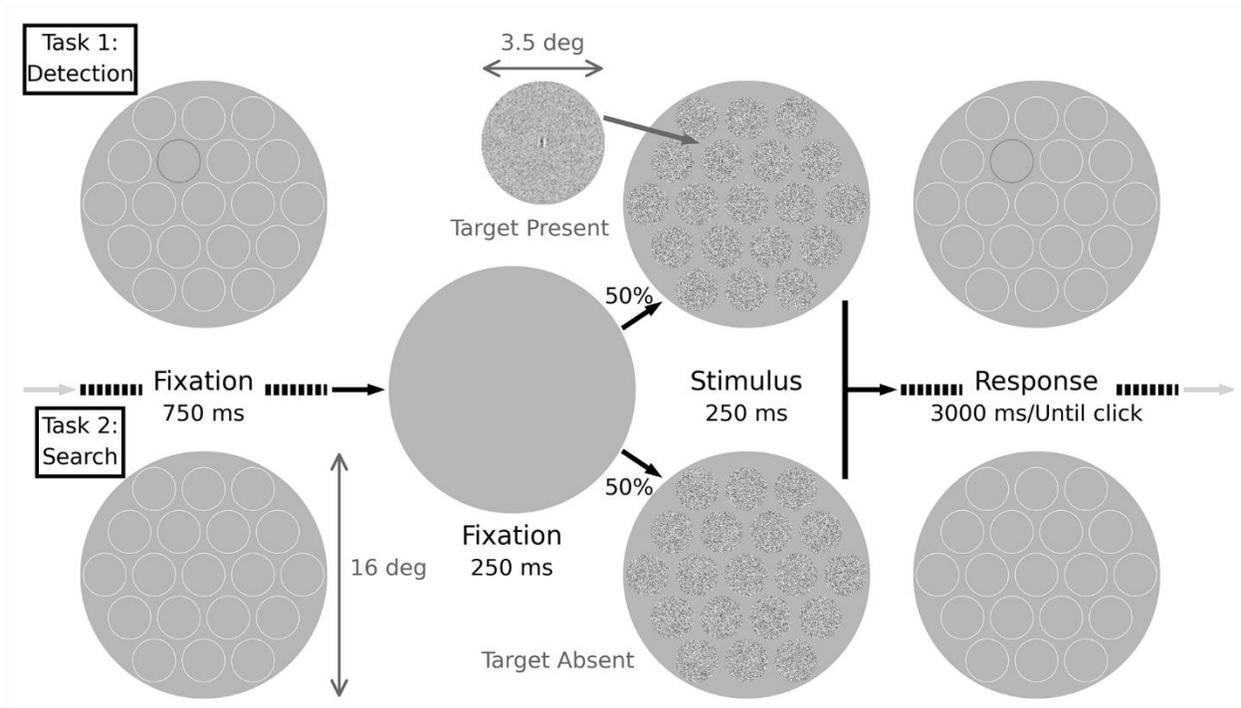

**Figure 1** Detection (Task 1) and search (Task 2) stimuli and timeline in a trial.

For the simple detection task, the observers were asked to indicate target absence with a right mouse click and target presence with a left mouse click. For search tasks, observers were asked to right click if the target was absent and to left click the judged location if present. We ran preliminary search trials with highly visible targets and found that the human observers made no errors in clicking on the target location, indicating that memory and motor-control limitations were not important factors in the study. Furthermore, we used a reverse counterbalancing design, where the observers completed the detection and search experiments in two opposite orders. This was done to keep any practice effects similar for the simple detection and search tasks.

For each observer, preliminary detection measurements were made at the central display location (fovea) to determine the target contrast giving a bias-corrected accuracy of approximately 95% correct ($d'$ = 4.5). This target contrast remained fixed for all search and detection trials.

The study included four male observers, aged 19–26. They all had normal or corrected-to-normal acuity. All procedures in the experiments were approved by the University of Texas Institutional Review Board.

## Results

The average target detectability ($d'$) at each of the 19 display locations (the $d'$ map) in the cued detection task is shown in Figure 2a. The average $d'$ across all locations is 2.17, and the over proportion correct is 84.2%. While there are some individual differences in these maps across the four observers, they show the same qualitative pattern: highest detectability in the fovea, intermediate at the 6 locations nearest the fovea, and poorest at the outer 12 locations, with relatively lower detectability in the upper and lower visual fields (Figure A1a-d). This qualitative pattern is consistent with previous studies (Carrasco et al., 1995; Carrasco & McElree, 2001; Najemnik & Geisler, 2005; Rovamo & Virsu, 1979).

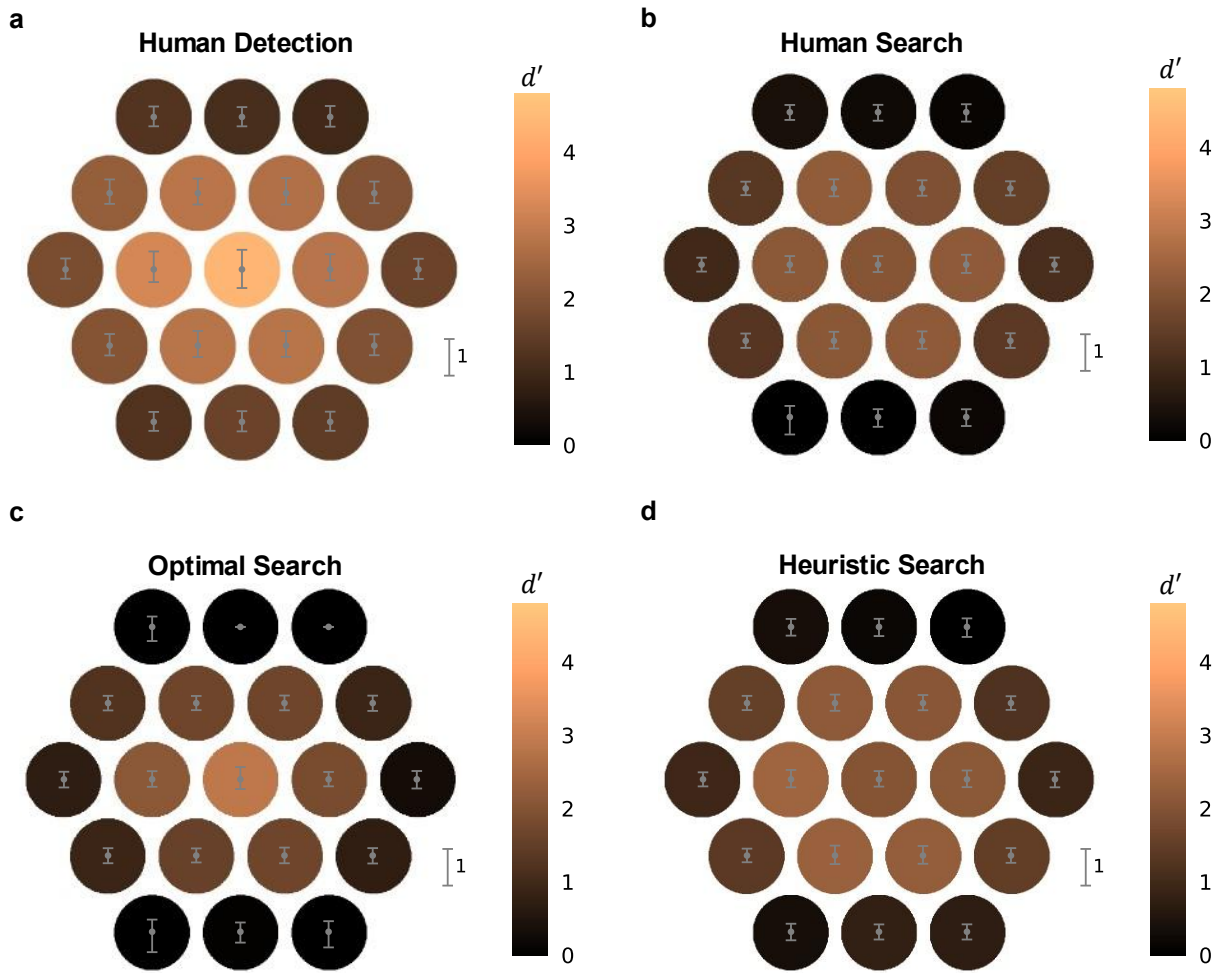

**Figure 2** Detection and search performance. **a**. Human detectability (d') map for the detection task. **b**. Human d' map for the search task. **c**. Optimal d' map for the search task. **d**. Heuristic d' map for search task, given correlated noise and foveal neglect.

The average detectability in the covert search task is shown in Figure 2b. Here, the detectability was computed from the hit rate at each target location and the overall correct rejection rate. The overall detectability is 1.22, corresponding to an overall accuracy of 69.8%, which is considerably lower than that in the detection task. This pattern is seen in all four observers (Appendix Table A1). Although there is a falloff in $d'$ with eccentricity, the $d'$ within the central 7 locations is relatively constant. This pattern holds for individual observers (Figure A1e-h).

The gray bars in Figure 3 show the average pattern of correct responses and errors, across observers, for the central location (Figure 3a), the surrounding 6 locations (Figure 3b), the outer 12 locations (Figure 3c) and the average across all locations (Figure 3d). Note that the false hit rate (FH) is the proportion of trials where the observer reported a location different from the target location.

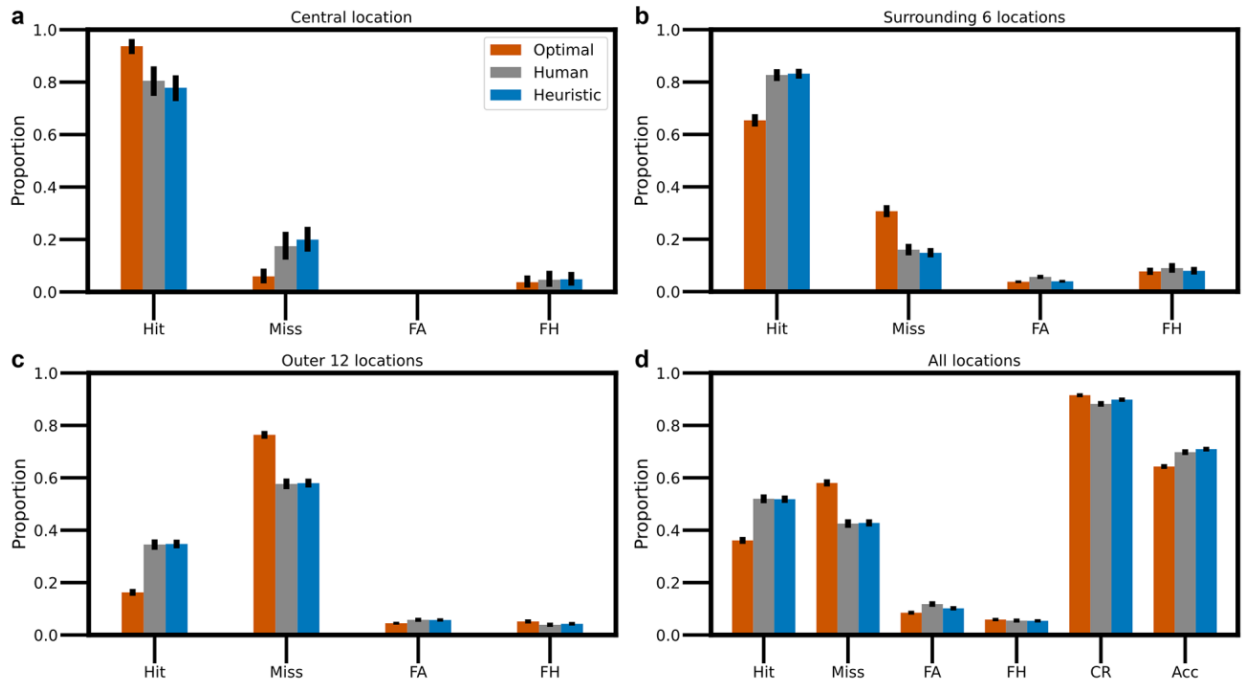

**Figure 3** Correct and incorrect responses in the search task by retinal eccentricity. **a**. Histograms of hits, misses, false alarms (FA) and false hits (FH) in the central location, for the average observer (gray), the optimal observer assuming statistical independence of responses across the 19 locations (orange), and a heuristic model observer (blue). **b**. Histograms for the surrounding six locations. **c**. Histograms for the outer twelve locations. **d**. Histograms for all locations; correct rejections and overall accuracy are also included. (Error bars are bootstrapped 95% confidence intervals.) The fit of the heuristic model observer is much better than the optimal searcher for the combined human observer and for all the individual human observers. The heuristic fit: log-likelihood = -11758, AIC = 23529. The optimal searcher fit: log-likelihood = -12039, AIC = 24078. The fall-off model is $e^{274.5}$ times as probable as the optimal searcher.

**Optimal observer for covert search**

To understand the relationship between the performance in the covert search and detection tasks, it is useful to consider what to expect from a covert searcher that uses the Bayes optimal decision rule. Given that we have directly measured the cued detectability of the target at each potential target location, this modeling analysis can be done within the standard signal detection theory framework (Green & Swets, 1966). We have shown (Oluk & Geisler, 2024; Walshe & Geisler, 2022; Zhang et al., 2023) that when the receptive field response is normally distributed, the Bayes optimal decision rule is given by

$$\hat{k} = \arg\max_{k \in [0,n]} \left( \ln p_k + d'_k R'_k - 0.5 d'^2_k \right) \qquad (1)$$

where $n$ is the number of potential target locations, $\hat{k}$ is the estimated target location, $p_k$ is the prior probability that the target is at location $k$, $d'_k$ is the detectability of the target at location $k$ in the cued detection task, and $R'_k$ is the normalized response on that trial at location $k$. In the standard signal detection framework, $R'_k$ is a random sample from a Gaussian distribution with a standard deviation of 1.0, and a mean of $d'_k$ when the target is present and a mean of 0.0 when the target is absent. In the signal detection framework, $R'_k$ represents the normalized log likelihood ratio of target present versus absent and can be thought of as the observer's decision variable in the cued detection task. Note that $k = 0$ represents the event that the target is absent (with $d'_0 = 0$). Note that if the search task is to only report target absent or present (not the location of the target when present), then the max rule in Equation 1 is not the optimal rule (Green & Swets, 1966; Burgess, 1985; Oluk & Geisler, 2024).

Figure 2c and the orange bars in Figure 3 show covert search performance using the optimal decision rule, given the average measured $d'$ map shown in Figure 2a, the uniform prior probabilities used in the experiment, and the assumption of statistical independence of the responses from the potential target locations. Statistical independence is plausible because the targets were small, the potential target locations were well separated, and the random noise backgrounds were statistically independent. While there is some general qualitative agreement between the Bayes optimal searcher and the average human searcher, there are several puzzling facts.

First, it is implausible that human observers implement calculations equivalent to the Bayes optimal decision rule. The optimal decision rule requires weighting the response at each potential target location by the detectability of the target at that location and adding the log of the prior probability of the target appearing at that location (see Equation 1). Learning all 19 detectabilities and 19 priors during the experiment seems unlikely. Worse yet, under natural conditions the $d'$ map is different on every fixation, even for the same target, because the masking properties of the background are different on every fixation. Also, the prior probability map varies depending on the scene context. For the optimal decision rule to be implemented under natural conditions, the human visual system would need extremely sophisticated neural mechanisms to estimate in parallel, during each fixation, the $d'$ map over the visual field for any desired target. Similarly, sophisticated mechanisms would be required to estimate the prior map from the current scene context. Thus, it is highly likely that the observers are using heuristic decision rules, making it unlikely that they can reach the performance predicted by the optimal decision rule.

Second, the overall accuracy of the average human observer is, in fact, slightly higher than that predicted by the optimal decision rule (see Figure 3d). This is also true for the individual human observers and for different numbers of potential target locations (Table A1).

Finally, although the overall accuracy of the human observers is higher than predicted by the optimal decision rule, their performance is suboptimal in the central location (compare Figures 2b and 2c and see Figure 3a). This result holds for all our observers and with varying numbers of potential target locations (Figures A1, A2, A6). This phenomenon, which we call "foveal neglect", replicates an earlier study measuring covert search performance in continuous noise backgrounds (Walshe & Geisler, 2022). In principle, foveal neglect should guarantee that human search performance falls below that predicted by the optimal decision rule.

What possible explanation is there for these seemingly paradoxical results? We argue that there are three factors that together could explain the results and that have broad implications for understanding and predicting search performance under natural conditions. First, and most important, a wide range of extremely simple heuristic decision rules can achieve optimal overall performance. Second, foveal neglect primarily affects only the central location out of the 19 locations. Third, correlated neural noise would cause the measured $d'$ values in the detection task to be an underestimate of the effective $d'$ values in the search task.

**Heuristics decision rules**

Consider first the effect of heuristic decision rules. A principled approach for evaluating heuristic decision rules is to quantitatively compare their performance with that of the Bayes optimal decision rule. To do this we simulated optimal and heuristic covert search performance for a wide range of possible $d'$ maps. Roughly speaking, detectability falls off with retinal eccentricity according to a two parameter function, where one parameter is the peak value of $d'$ in the center of the fovea, $d'_{max}$, and the other is the eccentricity at which $d'$ falls by a factor of 2 from the maximum, $e_2$ (see Equation 6 in the Appendix). For example, the average $d'$ map in Figure 2a is best fit with a $d'_{max}$ of 4.69 and an $e_2$ of 4.68. The colored curves in Figure 4a show cross sections of $d'$ maps with a $d'_{max}$ of 6.0, for $e_2$ values ranging from 1° to 9°. This range represents, approximately, the falloffs for wavelet targets from about 1 cycle/deg to 16 cycles/deg (Cannon, 1985; Peli et al., 1991), a range of spatial frequency that covers most of the area under the human contrast sensitivity function (CSF), and hence probably covers most search targets. Fine targets have small values of $e_2$, coarse targets have large values of $e_2$.

The solid black line in Figure 4b shows the overall accuracy in the covert search task of the Bayes optimal searcher for values of $e_2$ ranging from 1° to 9°, when the prior probability of target absent is 0.5 and the prior probability over all 19 target locations is uniform. The solid gray line shows the performance of the optimal searcher when the prior probability of target absent is 0.0 (i.e., a target is always present somewhere), and the prior probability over all 19 target locations is uniform. The optimal searcher always uses the exact $d'$ map for the target (see Equation 1). As expected, the accuracy of the optimal searcher increases substantially as the value of $e_2$ increases.

One of the simplest possible heuristic decision rules is to assume a completely flat $d'$ map: the horizontal dotted line in Figure 4a. The dotted curves in Figure 4b show the performance of a heuristic searcher that uses this single fixed flat $d'$ map. As can be seen, performance is nearly identical to that of the optimal decision rules. A slightly more complex but still easy to implement heuristic is a $d'$ map with a fixed peak and fixed falloff: the dashed curve in Figure 4a. The dashed curves in Figure 4b show the performance of this heuristic searcher, which is also near optimal.

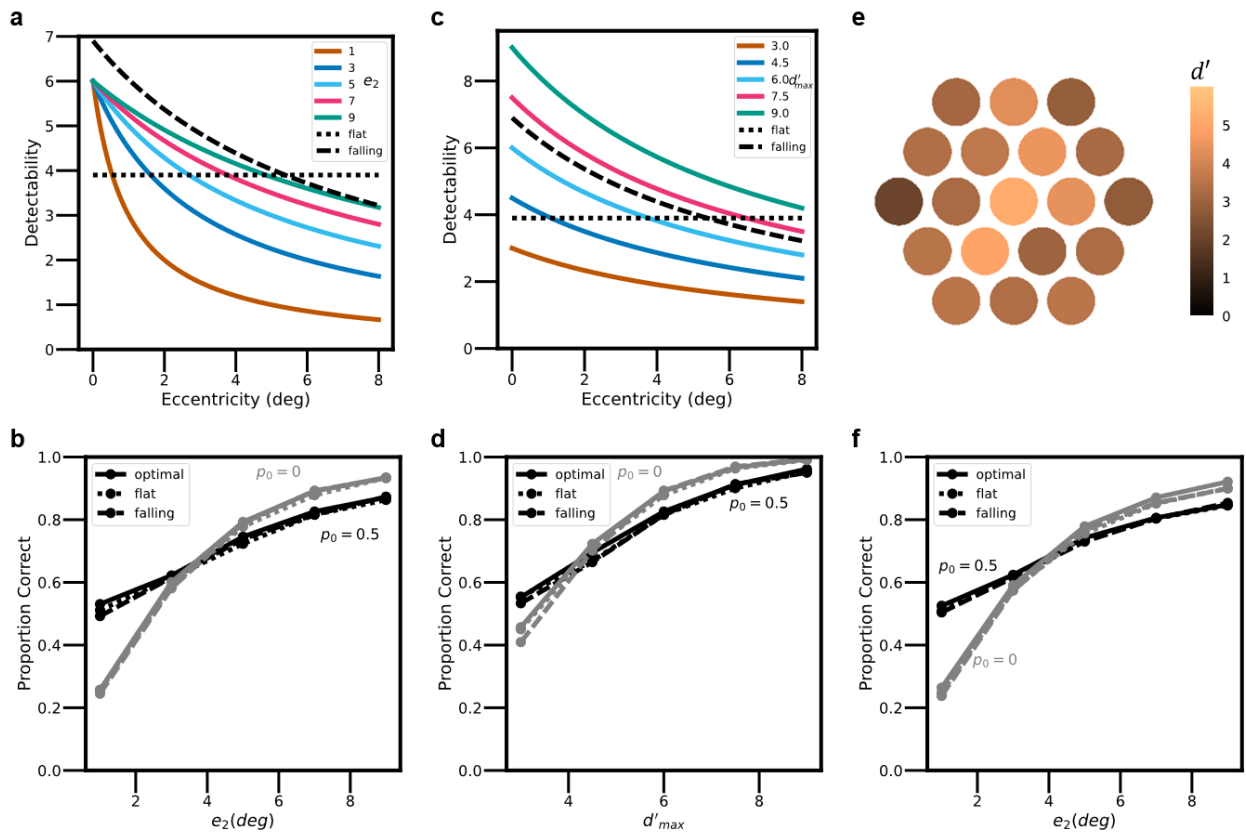

**Figure 4** Optimal versus heuristic decision rules. **a**. Actual maps for a range of $e_2$ values with a fixed $d'_{max}$ of 6.0 (colored curves). A flat heuristic map (dotted line). A heuristic map with the same fall-off as the best fit heuristic to the average human observer (dashed curve). **b**. Accuracy for optimal and heuristic searchers, for conditions in **a**. and target absence rate ($p_0$) of 0.5 and 0. **c**. Actual maps for a range of $d'_{max}$ values with a fixed $e_2$ of 7.0 (colored curves). The same flat heuristic map as in **a** (dotted line). The same heuristic map with a fall-off as in **a** (dashed curve). **d**. Accuracy for optimal and heuristic searchers for conditions in **c** and target absence rate ($p_0$) of 0.5 and 0. **e**. An example of the d' map varying randomly per trial. The baseline d' map has a $d'_{max}$ of 6.0 and a $e_2$ of 7.0. **f**. Accuracy for optimal and heuristic searchers for the baseline d' map with a $d'_{max}$ of 6.0 and $e_2$ ranging from 1° to 9°.

These results strongly suggest that the human visual system uses a highly heuristic rule, because there would be little or no benefit in implementing the much more complex optimal rule. If the visual system does use a heuristic rule, can that rule be estimated from behavioral data even though many heuristics give near optimal overall performance? The answer appears to be at least a partial yes. The dashed curve in Figure 4a is, in fact, the heuristic falloff parameter ($e_2 = 7°$) that best fits the pattern of corrects and errors (13 statistics in Figure 3: hit, miss, false hit, false alarm rates in the central, surrounding 6, and outer 12 locations, and the overall correct rejection rate). The blue histograms in Figure 3 show the predictions for this falloff parameter.

We also tested the same heuristic rules for $d'_{max}$ varying from 3.0 to 9.0, while holding $e_2$ at 7.0 (see Figure 4c). Figure 4d shows that again both heuristic decision rules achieve near optimal performance in all those conditions.

Under natural conditions, the properties of the background scene vary substantially over space and hence the $d'$ map is generally different with every new fixation. We simulated this situation by having a baseline $d'$ map with various $e_2$ parameter values and then assumed that the random background properties cause the actual $d'$ value at each location to be scaled up or down by a random percentage that is normally distributed, with a standard deviation of 20%. Figure 4e shows a single example of a random $d'$ map, where the baseline $e_2$ is 7.0 and the baseline $d'_{max}$ is 6.0. Figure 4f shows that the heuristic decision rules in Figures 4a achieve near optimal performance even when the $d'$ map varies randomly on each trial.

Similar results to those in Figure 4 hold for all combinations of $e_2$ and $d'_{max}$ (Figure A4), and for prior probability maps (Figure A9). Indeed, very simple heuristics are often near optimal under a very wide range of natural conditions. Nevertheless, heuristic rules with extreme deviations from the actual $d'$ and prior-probability maps are detrimental to search performance. For example, if there are substantial search regions where the $d'$ or prior probability go to zero, then no weight should be assigned to those locations in the decision. Similarly, performance can suffer substantially if regions where the $d'$ and prior probability are non-zero are given no weight in the decision.

**Foveal neglect**

Second, consider the phenomenon of foveal neglect. Even though humans perform slightly better than the optimal searcher, they have worse search performance in the foveal region (Figures 2b-c and 3a). In a previous study (Walshe & Geisler, 2022), we showed that this reduction in accuracy in the fovea is not due to bias in estimating the prior probability, but to a reduction in detectability in the fovea. We also showed that given a fixed total amount of attentional gain resources, foveal neglect is a principled phenomenon. The argument is that if there were limited attentional gain resources, they would be applied to some fixed number of cortical neurons, not to some fixed number of image pixels. Because of foveation, there are fewer neurons encoding the periphery and they have larger receptive fields. Thus, to maximize search performance over large regions of space it is better to assign less attentional gain to the dense, small-receptive-field neurons encoding the image near the line of sight.

To quantitatively evaluate this hypothesis, we assumed a fixed total amount of attention gain, distributed over a retinotopic map based on the measured anatomical density of the ganglion cells in the human retina (Drasdo et al., 2007; Watson, 2014). This retinotopic map is represented by the grid of lines in Figure 5a and is similar to the retinotopic map of primary visual cortex (Adams et al., 2007; Wässle et al., 1989). The color shading in Figure 5a and the curve in Figure 5b show the smooth modulation of attentional gain with retinal eccentricity (see Equation 3 in Methods) that best explains the data in Figures 2b and 3. As can be seen, attention gain is lowest in the fovea and increases to a maximum in the periphery.

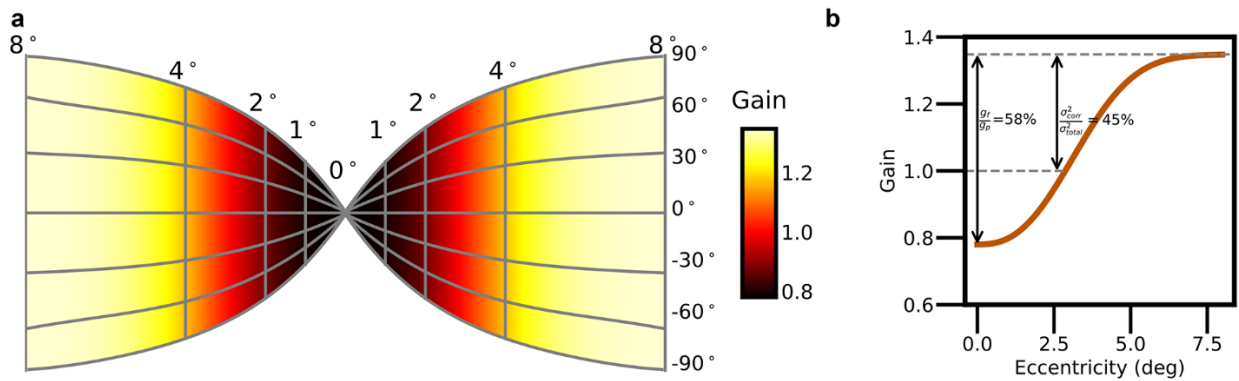

**Figure 5.** Foveal neglect and correlated noise. **a**. Retinotopic map. The flattened cortical sheet has a constant density of neurons. The grid of contours shows the retinal locations of the cortical neurons' receptive fields. **b**. The estimated variation in attentional gain with retinal eccentricity (also shown by coloring of map in **a**).

**Correlated noise**

Finally, consider the effects of correlated neural noise. Besides statistically independent sources of noise at each potential target location, we assume the existence of noise with common sources that are added to the responses at all potential target locations. These common sources cause the total noise at the different potential target locations to be partially correlated. For simplicity, we assume that the independent and common noise sources are Gaussian and statistically independent of each other, with standard deviations of $\sigma$ and $\sigma_0$, respectively. Thus, the total noise variance at each target location is $\sigma^2 + \sigma_0^2$. In the detection task, the correlated noise component necessarily lowers the detectability. However, the common noise has little or no effect on the optimal decision rule in the search task. For example, for a heuristic rule with a flat $d'$ map, the effect of the correlated noise on $d'$ is cancelled out by the max rule, because the correlated noise causes the same increase or decrease of responses at all potential target locations. Also, we note that the optimality of the max rule still holds even when the response is correlated across locations (Oluk & Geisler, 2024). We show in the Appendix that the effective $d'$ map for the search task compared to the detection task is scaled up by a factor of $\sqrt{\sigma^2 + \sigma_0^2}/\sigma$, which is equivalent to scaling by $1/\sqrt{1-r}$, where $r$ is the noise correlation (Abbey & Eckstein, 2000).

Typically, the attention gain cannot exceed 1.0, because the highest possible value of $d'$ should occur when the target location is cued. However, correlated noise allows the gain to exceed 1.0 in our search task. This happens because correlated noise reduces sensitivity in the cued detection task but not in the search task. The amount that the gain exceeds 1.0 provides an estimate of the fraction of the total noise variance that is due to correlated noise (see Equation 4 in the Appendix). In the present case, a correlated noise variance that is approximately 45% of the total noise variance explains the better search performance of human observers compared to the optimal searcher using the $d'$ map from the detection task (Figure 5). Similar estimates of foveal neglect and correlated noise were obtained for all four observers, although there are individual differences (see Figure A3).

In sum, the combination of the three factors described here provides a plausible, quantitatively accurate explanation of the seeming paradoxical detection and search results shown in Figures 2 and 3. The quantitative predictions of the best fitting heuristic searcher with these three factors are much more consistent with the human search behavior than the optimal searcher without these three factors.

## Discussion

Covert search and cued detection performance were measured for wavelet targets in Gaussian noise under carefully controlled conditions. The detectability ($d'$) map measured in the cued detection task was used to predict covert search performance for the Bayes optimal decision rule, assuming statistically independent sensory responses from the potential target locations. We found that human performance slightly exceeded the predictions of the Bayes optimal decision rule, despite the complexity of the optimal rule, and despite the fact the humans showed a loss of sensitivity in the fovea (foveal neglect). We found that these seeming paradoxical results can be explained quantitatively by three facts: (1) very simple heuristic decision rules, together with local normalization, can achieve essentially optimal performance, (2) foveal neglect primarily reduces the effective $d'$ value at the central target location, and (3) correlated neural noise causes the $d'$ values measured in the detection task to underestimate the effective $d'$ values in the search task.

### Generality

An obvious question is how general these findings are. We have also made some measurements of covert search for 7, 61 and 91 potential target locations (Figure A2, Table A1). The results are generally consistent with those for 19 locations shown in Figures 2 and 3. However, we are less confident about the model predictions for 61 and 91 locations, because the $d'$ maps needed to be interpolated and extrapolated from the 19 locations. Importantly, we find that the heuristic decision rules work equally well for 7, 19, 61, and 91 locations. Thus, it seems safe to conclude that the findings will hold over a wide range of potential target locations.

We chose a 6-cpd wavelet target because the detectability of this target varies substantially over the 16º diameter search region. It would be informative to repeat the measurements for other targets, but given what is known from the detection and search literature, and the fact the simple heuristic decision rules are near optimal for a wide range of $d'$ maps, it is likely that the findings will generalize well across a wide range of spatially-localized targets.

We chose white-noise backgrounds because they have a dense naturalistic structure yet are statistically simple, and because they have been widely used in studies of visual performance. Natural backgrounds, on the other hand, are statistically complex and non-stationary, so that the masking properties of the background typically vary across potential target locations. Thus, the $d'$ map on each trial (e.g., eye fixation) is due to a combination of foveation and variation in the masking properties of the background. The fact that simple fixed heuristic decision rules are effectively optimal even when the $d'$ map changes randomly on each trial (Figures 4e–f, and Figure A7) suggests that our findings may hold across a wide range of stationary and non-stationary backgrounds. The difficulty in directly testing this hypothesis is that to calculate optimal performance one must know the $d'$ map for the stimulus on each trial. Estimating the $d'$ map is tractable for white noise backgrounds with luminance and contrast varying across spatial locations, because of the lawfulness of the effects of noise contrast and luminance (Burgess et al., 1981; Lu & Dosher, 1999; Sebastian et al., 2017), but is more problematic for natural backgrounds. However, progress has been made in modeling and predicting detectability in arbitrary natural backgrounds (Sebastian et al., 2017, 2020; Walshe et al., 2020; Zhang et al., 2023; Zhang & Geisler, 2022). Given the current study, it is possible that such models will be able to predict human covert search performance in natural backgrounds by including simple heuristic decision rules, correlated noise, and foveal neglect.

Our theoretical analysis and modeling make the standard signal-detection-theory assumption that the observer's decision variable in the detection task is normally distributed. The psychophysical literature shows that this simplifying assumption is robust. Also, in previous work we have shown that linear receptive field responses to natural backgrounds are approximately normally distributed if the responses are normalized by the values of the background properties (e.g., luminance and contrast) that are known to have a substantial masking effect (Sebastian et al., 2017); this finding is also consistent with earlier related work (Schwartz & Simoncelli, 2001).

Where humans tend to fall below the predictions of the optimal decision rule is when the task is to identify the locations of multiple targets (Verghese, 2012), identify the target location when the target may be drawn many different categories of target or, more generally, when there are substantial demands on memory or on high-level cognitive computation (e.g., which search location contains a number divisible by 13). Theories of covert search for such conditions must include limitations in memory and cognitive computation. Nonetheless, there are many real-world situations where observers are covertly searching for targets that require low cognitive effort.

Our findings for the covert search task are likely to generalize to many other identification tasks, such as facial recognition. Most identification tasks can be described as choices between mutually exclusive events. For such tasks the max rule remains optimal or near optimal. Thus, a Bayesian heuristic decision analysis like the one described here may provide useful insights and testable predictions.

**Normalization**

Normalization is a fundamental and ubiquitous property of cortical processing (Albrecht & Geisler, 1991; Carandini & Heeger, 2012; Heeger, 1991). The present results suggest that it may play a more important role in perceptual decision making than previously appreciated. For detection tasks in natural backgrounds, it has been shown that normalization by local luminance, contrast, and similarity allows near-optimal decisions with a single fixed decision criterion (Sebastian et al., 2017). The present results greatly expand this conclusion by showing that such normalization should also allow extremely simple heuristic decision rules to achieve near optimal performance for a wide range of natural identification tasks. Without normalization, the heuristics described here do not work nearly as well on natural and other non-stationary backgrounds. For example, if the detectability at different locations varies because of differences in the contrast of the white noise, then simple fixed heuristics without normalization give substantially poorer performance (see Figure A8).

**Evolution and individual differences**

The existence of simple fixed decision rules that are near optimal increases the likelihood of developing accurate quantitative theories of covert search under natural conditions. It is relatively easier for evolution to find such decision rules, so there is higher likelihood that near-optimal rules are implemented in real visual systems. Also, the simplicity of these near-optimal rules should make it easier to uncover the neural circuits that implement them.

An important implication of our heuristic decision analysis is that there can be substantial individual differences in covert-search decision rules with no effect on overall search accuracy. In other words, seeing individual differences in the patterns of correct responses and errors, or in the trial-by-trial responses to the same stimuli, does not imply that one of the observers has a better search strategy.

Another implication is that the neural circuitry for optimal covert-search decisions is likely to be relatively simple. Thus, animals with relatively few neural resources should be able to approximate the same optimal decision rules as well as animals with many more neural resources.

Finally, the fact that many simple fixed heuristics perform optimally may have important implications for the evolution of decision processes. Specifically, natural selection may cause an increase in individual differences in covert-search decision rules. This may be especially true in social species, because bigger individual differences may reduce competition in foraging and increase the probability that the group detects predators, prey, and other targets of value to the group. At the same time, this increase in individual differences would not be detrimental to the individual when away from the group.

**Prior probabilities**

The optimal Bayesian decision rule for covert search also takes into account the prior probability map (see Equation 1). In the current experiments, the prior probability of target absent was 0.5 and prior probability across the potential target locations was uniform. Given the simplicity of this prior map, we did not focus on the effect of assuming simple heuristics. However, in Figure A9, we report some initial simulations of simple heuristics (uniform priors) when the actual prior maps are not uniform. We find that it is important to know the target-absent prior probability reasonably well in order to approach optimal performance. However, as with the $d'$ maps, a uniform target-present prior map achieves near optimal overall performance for a wide range of actual prior maps.

**Bayesian decision making**

The classic view in the Bayesian decision making literature is that taking into account reliabilities and prior probabilities is critical for making decisions that optimize accuracy (Geisler, 2011; Knill & Richards, 1996; Ma et al., 2022). Our simulations and experimental results show that this is not true for covert visual search tasks and probably many other identification tasks (Oluk & Geisler, 2024). While it is important to take into account the regions of stimulus space where reliabilities and priors go essentially to zero, it is not important to take into account the specific pattern of values within the non-zero regions. This finding depends on the assumption of parallel processing, which holds when applying the max rule. Parallel processing also holds for most deep neural networks (DNNs) and hence DNNs also benefit from the fact that they can achieve near optimal overall performance without taking into account detailed variations in prior probability and reliability. It follows that DNNs can differ in their responses to specific stimuli even when their overall performance is similar and near optimal. If processing is more limited and serial, then effective decision rules need to take into account detailed priors and reliabilities. For example, in a covert search task with a 50% prior of target absent it would be a mistake to spend the limited amount of processing time on a few locations with low prior probability and/or low reliability.

**Correlated noise**

The max rule is the optimal decision rule not just for covert search, but for many identification tasks. For these tasks, correlated noise (across all the category responses) has a negligible or minor effect on the accuracy of the decision. We showed that correlated noise can create a mismatch between the $d'$ values estimated in the cued detection task and the effective $d'$ values in the covert search task. Such mismatches might occur in other identification tasks, and thus would be an important factor to consider.

Our results do not prove that correlated noise is the source of the supraoptimal accuracy of the observers in our experiments, but it is a plausible hypothesis because there are many likely sources of correlated noise, such as slow modulations in membrane potential and in more indirect neural-response measures such as the bold response (Cohen & Kohn, 2011; Cowley et al., 2020; Kohn et al., 2009, 2016; Rosenbaum et al., 2017).

An interesting possibility is that the nervous system evolved to inject correlated variations into the pathways transmitting information to the circuits or brain areas that perform identification tasks. Because these correlated variations do not hurt identification performance, they could provide an independent (low bitrate) channel for communicating other kinds of information such as reward signals, arousal signals, task-relevant global context information, etc. Such correlated variations could be used to broadcast these other kinds of information to any of the decision-making circuits that use the max rule. Piggybacking on existing major pathways could also reduce the need for separate specialized minor pathways. The benefits of this low bit rate communication channel may outweigh the cost of reduced sensitivity in cued detection tasks. A recent study provides evidence for the related hypothesis that stochastic co-modulation of specific neural populations serves as a label for task relevance in subsequent stages of processing (Haimerl et al., 2023).

**Intrinsic position uncertainty**

In a previous study of covert search in backgrounds with pixel-wise target locations effectively continuous to the human visual system (Walshe & Geisler, 2022), we showed that it is important correct for the effects of intrinsic position uncertainty when comparing cued detection performance with covert search performance for the same targets and backgrounds. Not correcting for intrinsic position uncertainty with continuous backgrounds leads to predictions of optimal search performance that fall below human search performance. This occurs because the extrinsic uncertainty subsumes the intrinsic uncertainty and hence not correcting for intrinsic uncertainty over weights (double counts) the effect of intrinsic uncertainty in the covert search task. This effect of intrinsic uncertainty does not apply in the current study because the intrinsic uncertainty is contained within the discrete background patches and the extrinsic uncertainty is only across the background patches (the extrinsic uncertainty does not subsume the intrinsic uncertainty).

**Overt search**

Visual search under natural conditions is a mixture of covert and overt search. During the first part of each fixation there is a covert search event to identify the location of the target or potential target locations. During the second part of each fixation the next fixation location is computed. The optimal decision rule for picking the next fixation location also takes into account the $d'$ map and the prior probability map (Najemnik & Geisler, 2005). An important next step will be to perform a Bayesian heuristic decision analysis for fixation selection.

## Data availability

All the data that support the findings of this study are available via GitHub at https://github.com/anqi-j/supraoptimal-covert-search.

## Code availability

All codes that analyze the findings of this study are available via GitHub at https://github.com/anqi-j/supraoptimal-covert-search.

## Acknowledgements

Supported by NIH grants EY024662 and EY11747.

## References

Abbey, C. K., & Eckstein, M. P. (2000). Derivation of a detectability index for correlated

responses in multiple-alternative forced-choice experiments. *JOSA A, Vol. 17, Issue 11,*

*Pp. 2101-2104*. https://doi.org/10.1364/JOSAA.17.002101

Adams, D. L., Sincich, L. C., & Horton, J. C. (2007). Complete Pattern of Ocular Dominance

Columns in Human Primary Visual Cortex. *Journal of Neuroscience*, *27*(39), 10391–

10403. https://doi.org/10.1523/JNEUROSCI.2923-07.2007

Albrecht, D. G., & Geisler, W. S. (1991). Motion selectivity and the contrast-response function of simple cells in the visual cortex. *Visual Neuroscience*, *7*(6), 531–546.

Bochud, F. O., Abbey, C. K., & Eckstein, M. P. (2004). Search for lesions in mammograms:

Statistical characterization of observer responses. *Medical Physics*, *31*(1), 24–36.

https://doi.org/10.1118/1.1630493

Brainard, D. H. (1997). The Psychophysics Toolbox. *Spatial Vision*, *10*(4), 433–436.

https://doi.org/10.1163/156856897X00357

Burgess, A. (1985). Visual signal detection.III.On Bayesian use of prior knowledge and cross correlation. *JOSA A, Vol. 2, Issue 9, Pp. 1498-1507*. https://doi.org/10.1364/JOSAA.2.001498

Burgess, & Ghandeharian. (1984). Visual signal detection. II. Signal-location identification. *JOSA A, Vol. 1, Issue 8, Pp. 906-910*. https://doi.org/10.1364/JOSAA.1.000906

Burgess, Wagner, Jennings, & Barlow. (1981). Efficiency of Human Visual Signal Discrimination. *Science*, *214*(4516), 93–94. https://doi.org/10.1126/science.7280685

Cannon, M. W. (1985, October 14). Contrast perception in the peripheral visual field. *Annual Meeting Optical Society of America (1985), Paper WJ41*. https://doi.org/10.1364/OAM.1985.WJ41

Carandini, M., & Heeger, D. J. (2012). Normalization as a canonical neural computation. *Nature Reviews Neuroscience*, *13*(1), 51–62. https://doi.org/10.1038/nrn3136

Carrasco, M. (2011). Visual attention: The past 25 years. *Vision Research*, *51*(13), 1484–1525. https://doi.org/10.1016/j.visres.2011.04.012

Carrasco, M., Evert, D. L., Chang, I., & Katz, S. M. (1995). The eccentricity effect: Target eccentricity affects performance on conjunction searches. *Perception & Psychophysics*, *57*(8), Article 8. https://doi.org/10.3758/BF03208380

Carrasco, M., & McElree, B. (2001). Covert attention accelerates the rate of visual information processing. *Proceedings of the National Academy of Sciences*, *98*(9), 5363–5367. https://doi.org/10.1073/pnas.081074098

Cohen, M. R., & Kohn, A. (2011). Measuring and interpreting neuronal correlations. *Nature Neuroscience*, *14*(7), 811–819. https://doi.org/10.1038/nn.2842

Cowley, B. R., Snyder, A. C., Acar, K., Williamson, R. C., Yu, B. M., & Smith, M. A. (2020). Slow

   Drift of Neural Activity as a Signature of Impulsivity in Macaque Visual and Prefrontal

   Cortex. *Neuron*. https://doi.org/10.1016/j.neuron.2020.07.021

Drasdo, N., Millican, C. L., Katholi, C. R., & Curcio, C. A. (2007). The length of Henle fibers in the

   human retina and a model of ganglion receptive field density in the visual field. *Vision

   Research*, *47*(22), 2901–2911. https://doi.org/10.1016/j.visres.2007.01.007

Eckstein, M. P. (1998). The Lower Visual Search Efficiency for Conjunctions Is Due to Noise and

   not Serial Attentional Processing. *Psychological Science*, *9*(2), 111–118.

   https://doi.org/10.1111/1467-9280.00020

Eckstein, M. P. (2011). Visual search: A retrospective. *Journal of Vision*, *11*(5), 14–14.

   https://doi.org/10.1167/11.5.14

Eckstein, M. P., Thomas, J. P., Palmer, J., & Shimozaki, S. S. (2000). A signal detection model

   predicts the effects of set size on visual search accuracy for feature, conjunction, triple

   conjunction, and disjunction displays. *Perception & Psychophysics*, *62*(3), 425–451.

   https://doi.org/10.3758/BF03212096

Geisler, W. S. (2011). Contributions of ideal observer theory to vision research. *Vision Research*,

   *51*(7), 771–781. https://doi.org/10.1016/j.visres.2010.09.027

Green, D. M., & Swets, J. A. (1966). *Signal detection theory and psychophysics* (pp. xi, 455). John

   Wiley.

Greene, H. H. (2006). The Control of Fixation Duration in Visual Search. *Perception*, *35*(3), 303–

   315. https://doi.org/10.1068/p5329

Haimerl, C., Ruff, D. A., Cohen, M. R., Savin, C., & Simoncelli, E. P. (2023). Targeted V1 comodulation supports task-adaptive sensory decisions. *Nature Communications*, *14*(1), 7879. https://doi.org/10.1038/s41467-023-43432-7

Heeger, D. J. (1991). Nonlinear model of neural responses in cat visual cortex. In *Computational models of visual processing* (pp. 119–133). The MIT Press.

Hooge, I. Th. C., & Erkelens, C. J. (1998). Adjustment of fixation duration in visual search. *Vision Research*, *38*(9), 1295-IN4. https://doi.org/10.1016/S0042-6989(97)00287-3

Knill, D. C., & Richards, W. (Eds.). (1996). *Perception as Bayesian Inference*. Cambridge University Press. https://doi.org/10.1017/CBO9780511984037

Kohn, A., Coen-Cagli, R., Kanitscheider, I., & Pouget, A. (2016). Correlations and Neuronal Population Information. *Annual Review of Neuroscience*, *39*, 237–256. https://doi.org/10.1146/annurev-neuro-070815-013851

Kohn, A., Zandvakili, A., & Smith, M. A. (2009). Correlations and brain states: From electrophysiology to functional imaging. *Current Opinion in Neurobiology*, *19*(4), 434–438. https://doi.org/10.1016/j.conb.2009.06.007

Lu, Z.-L., & Dosher, B. A. (1999). Characterizing human perceptual inefficiencies with equivalent internal noise. *JOSA A, Vol. 16, Issue 3, Pp. 764-778*. https://doi.org/10.1364/JOSAA.16.000764

Ma, W. J., Kording, K. P., & Goldreich, D. (2022). *Bayesian models of perception and action*. Cambridge, MA: MIT Press.

Najemnik, J., & Geisler, W. S. (2005). Optimal eye movement strategies in visual search. *Nature*, *434*(7031), 387–391. https://doi.org/10.1038/nature03390

Oluk, C., & Geisler, W. S. (2024). Target identification under high levels of amplitude, size, orientation and background uncertainty. *bioRxiv : The Preprint Server for Biology*. https://doi.org/10.1101/2024.08.30.610264

Palmer, J., Ames, C. T., & Lindsey, D. T. (1993). Measuring the effect of attention on simple visual search. *Journal of Experimental Psychology: Human Perception and Performance*, *19*(1), 108–130. https://doi.org/10.1037/0096-1523.19.1.108

Palmer, J., & Davis, E. (2004). Visual search and attention: An overview. *Spatial Vision*, *17*(4–5), 249–255. https://doi.org/10.1163/1568568041920168

Peli, E., Yang, J., & Goldstein, R. B. (1991). Image invariance with changes in size: The role of peripheral contrast thresholds. *JOSA A, Vol. 8, Issue 11, Pp. 1762-1774*. https://doi.org/10.1364/JOSAA.8.001762

Pelli, D. G. (1997). The VideoToolbox software for visual psychophysics: Transforming numbers into movies. *Spatial Vision*, *10*(4), 437–442. https://doi.org/10.1163/156856897X00366

Rosenbaum, R., Smith, M. A., Kohn, A., Rubin, J. E., & Doiron, B. (2017). The spatial structure of correlated neuronal variability. *Nature Neuroscience*, *20*(1), 107–114. https://doi.org/10.1038/nn.4433

Rovamo, J., & Virsu, V. (1979). An estimation and application of the human cortical magnification factor. *Experimental Brain Research*, *37*(3), Article 3. https://doi.org/10.1007/BF00236819

Schwartz, O., & Simoncelli, E. P. (2001). Natural signal statistics and sensory gain control. *Nature Neuroscience*, *4*(8), 819–825. https://doi.org/10.1038/90526

Sebastian, S., Sebastian, S., Abrams, J., Abrams, J., Geisler, W. S., & Geisler, W. S. (2017). Constrained sampling experiments reveal principles of detection in natural scenes. *Proceedings of the National Academy of Sciences of the United States of America*. https://doi.org/10.1073/pnas.1619487114

Sebastian, S., Sebastian, S., Seemiller, E., Seemiller, E., Geisler, W. S., & Geisler, W. S. (2020). Local reliability weighting explains identification of partially masked objects in natural images. *Proceedings of the National Academy of Sciences of the United States of America*. https://doi.org/10.1073/pnas.1912331117

Verghese, P. (2012). Active search for multiple targets is inefficient. *Vision Research*, *74*, 61–71. https://doi.org/10.1016/j.visres.2012.08.008

Walshe, R. C., & Geisler, W. S. (2022). Efficient allocation of attentional sensitivity gain in visual cortex reduces foveal sensitivity in visual search. *Current Biology*, *32*(1), 26-36.e6. https://doi.org/10.1016/j.cub.2021.10.011

Walshe, R. C., Walshe, R. C., Geisler, W. S., & Geisler, W. S. (2020). Detection of occluding targets in natural backgrounds. *Journal of Vision*. https://doi.org/10.1167/jov.20.13.14

Wässle, H., Grünert, U., Röhrenbeck, J., & Boycott, B. B. (1989). Cortical magnification factor and the ganglion cell density of the primate retina. *Nature*, *341*(6243), 643–646. https://doi.org/10.1038/341643a0

Watson, A. B. (2014). A formula for human retinal ganglion cell receptive field density as a function of visual field location. *Journal of Vision*, *14*(7), 15. https://doi.org/10.1167/14.7.15

Zhang, A., & Geisler, W. S. (2022). Detection of targets in filtered noise: Whitening in space and spatial frequency. *JOSA A*, *39*(4), 690–701. https://doi.org/10.1364/JOSAA.447391

Zhang, A., Seemiller, E. S., & Geisler, W. S. (2023). Phase-dependent asymmetry of pattern masking in natural images explained by intrinsic position uncertainty. *Journal of Vision*, *23*(10), 16. https://doi.org/10.1167/jov.23.10.16

# Appendix

## Individual observers

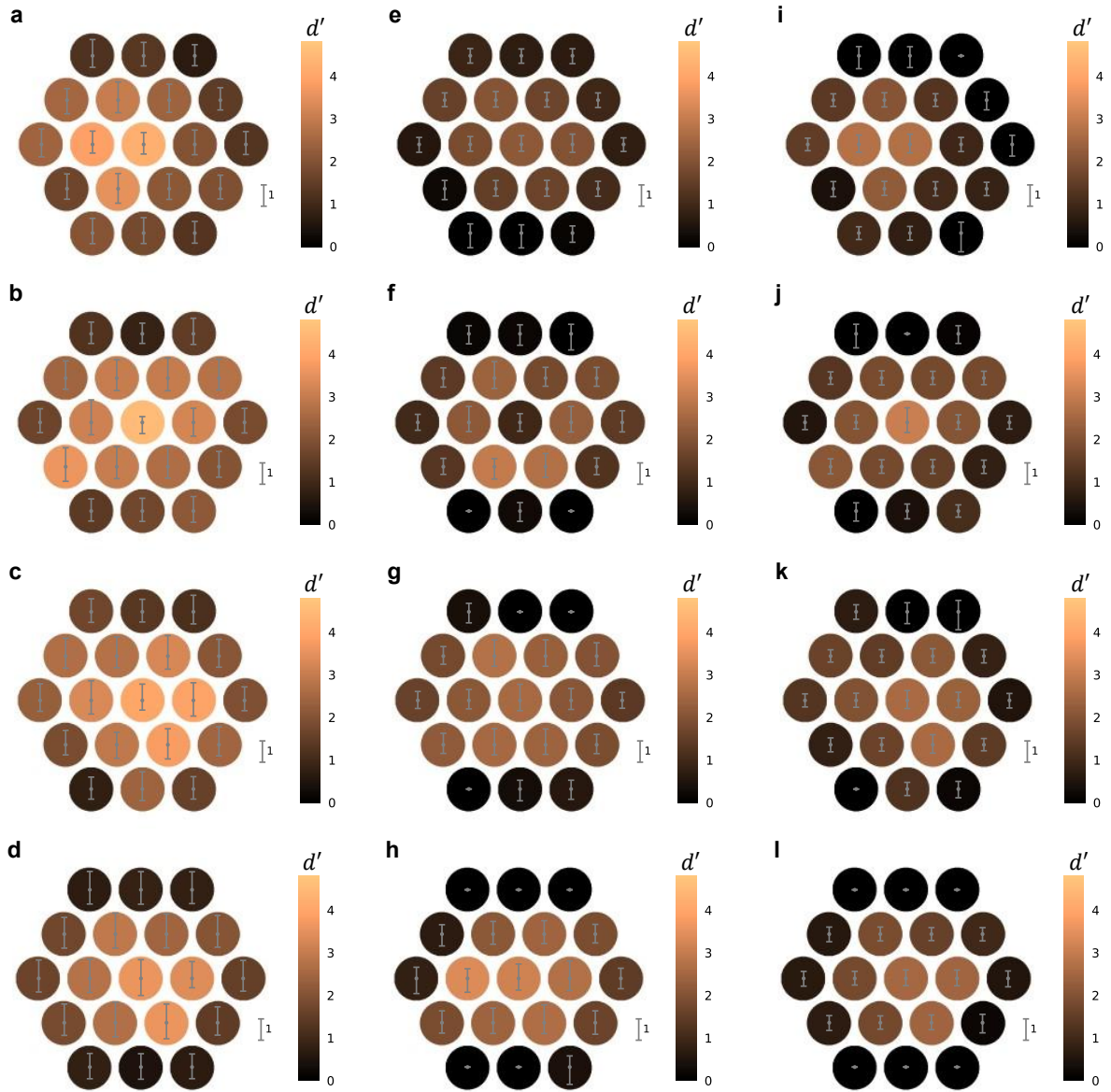

**Figure A1** Detectability maps per observer. **a-d.** In the detection experiment. **e-h.** In the search experiment. **i-l.** The optimal searcher in the search experiment, based on the $d'$ map per observer.

**Number of potential target locations**

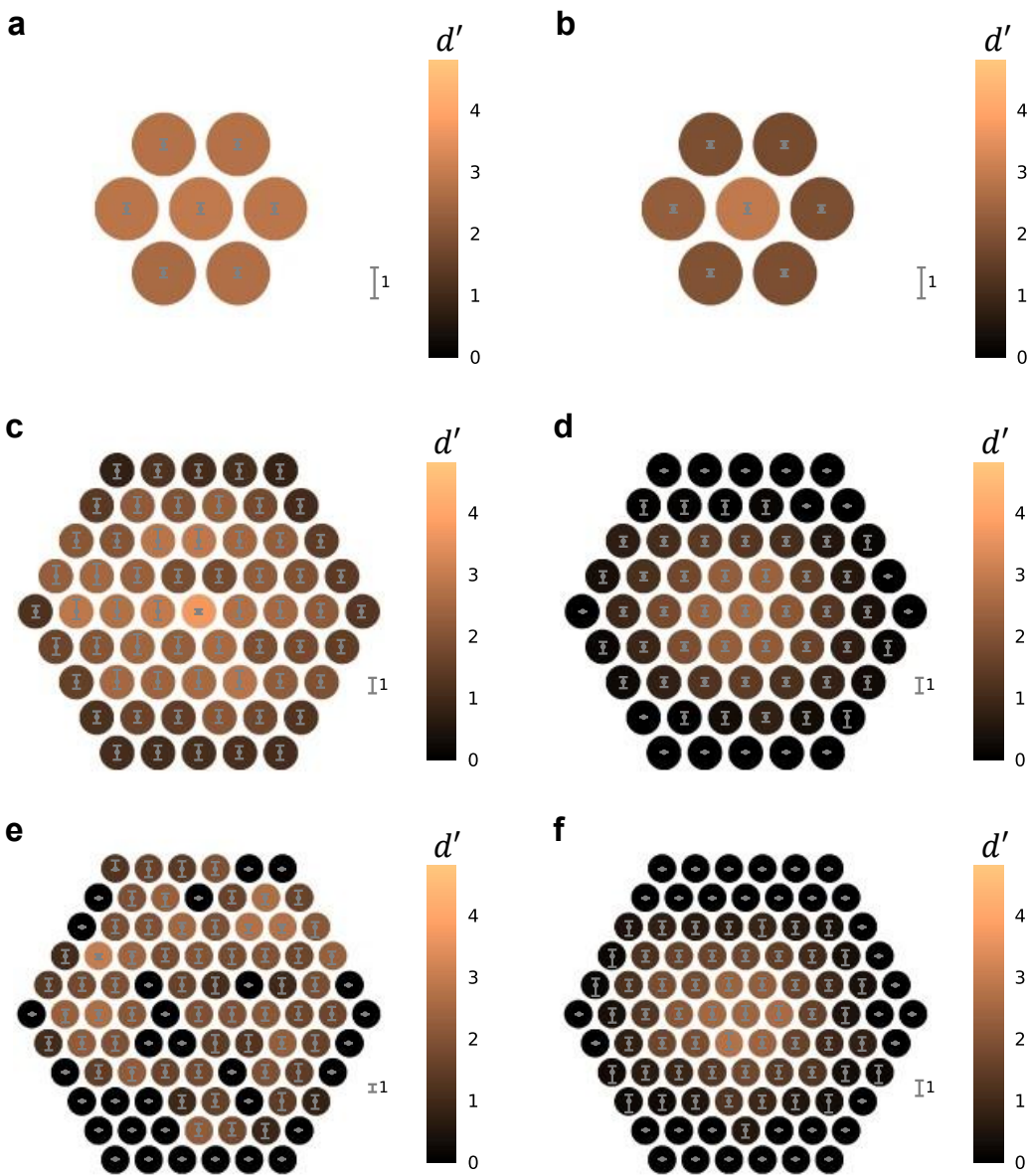

**Figure A2** Search detectability maps for varying numbers of search locations. **a**. Search performance of the average human observer when cued for the 7 central locations of the main search task. **b**. Search performance of an optimal searcher having the $d'$ map for the central 7 locations of the 19-location $d'$ map shown in Figure 1b. **c**. Search performance of the average human observer for 61 potential target locations. Target locations still cover the central 16 deg of the visual field. **d**. Search performance of the optimal searcher for 61 locations, using a $d'$ map interpolated and extrapolated from the 19-location $d'$ map shown in Figure 1b. **e**. Search performance of one human observer for 91 potential target locations. Target locations still cover the central 16 deg of the visual field. **f**.

Search performance of the optimal searcher for 91 locations, using a $d'$ map interpolated and extrapolated from the 19-location $d'$ map shown in Figure 1b.

| Task | No. of locations | Human/ Optimal Searcher | O1 | O2 | O3 | O4 | Average |
|---|---|---|---|---|---|---|---|
| Detection | | | 0.8167 | 0.8596 | 0.8522 | 0.7862 | 0.8325 |
| Search | 19 | Human | 0.6762 *** | 0.7069 *** | 0.7225 *** | 0.6958 *** | 0.6978 *** |
| | | Optimal | 0.6391 | 0.6841 | 0.6869 | 0.6237 | 0.6430 |
| | 7 | Human | 0.8979 *** | 0.9156 *** | 0.9512 *** | 0.8917 *** | 0.9135 *** |
| | | Optimal | 0.8193 | 0.8561 | 0.8871 | 0.8417 | 0.8438 |
| | 61 | Human | 0.7025 *** | 0.8319 *** | 0.8388 *** | / | 0.791 *** |
| | | Optimal | 0.5883 | 0.6316 | 0.6426 | / | 0.5978 |
| | 91 | Human | 0.7331 *** | / | / | / | 0.7331 *** |
| | | Optimal | 0.5873 | / | / | / | 0.5873 |

**Table A1** Proportion correct in all experiments for 4 human observers. The optimal-searcher accuracy was computed assuming the corresponding individual or combined $d'$ map. Asterisks indicate the p-values of the human accuracy to the accuracy distribution of the corresponding optimal searcher, that *: p-value < 0.05; **: p-value < 0.01; ***: p-value < 0.00001.

**Attentional gain and correlated noise**

The allocation of attentional sensitivity gain was modeled as in the efficiency-limited-foveated (ELF) searcher and the retinotopic map definitions in (Walshe & Geisler, 2022). The attentional efficiency gain $g(\mathbf{x})$ was applied to obtain a modified $d'$ map:

$$d'_g(\mathbf{x}) = g(\mathbf{x})d'(\mathbf{x}) \quad (2)$$

A retinotopic map is shown in Figure 4 and A3a. Figure A3a and A3b illustrates the cortical area corresponding to each of the 19 background regions. The attentional sensitivity gain map was modeled as a Weibull function

$$g(\mathbf{i}) = g_f + (g_p - g_f)\left[1 - e^{-(i/a)^b}\right] \quad (3)$$

where $\mathbf{i} = (i, j)$ is a coordinate in V1, $i$ represents the coordinate along the horizontal meridian, $g_f$, $g_p$ are the gains in fovea and periphery, respectively, and $a$, $b$ are the steepness and shape parameters of the gain modulation, respectively (for more details see (Walshe & Geisler, 2022)).

Spatially correlated Internal noise is another component of our model. We assume that the independent noise variance at each location is $\sigma^2$, the correlated noise added all the locations is $\sigma_0^2$, that the independent and correlated noise sources are independent of each other. Thus, in the detection experiment the detectability is $d'_d = a/\sqrt{\sigma_0^2 + \sigma^2}$ and the detectability in the search task is effectively $d'_s = a/\sigma$, because the correlated noise is cancelled by applying the max rule.

Simulations show that, for moderate levels of correlated noise, if one applies the max rule in the cued detection task by comparing the cued location to another location having equal detectability there is no improvement in performance. This occurs because the comparison turns the task into a same-different discrimination task, which has square-root of two lower detectability than a yes-no task[14].

Because of the correlated noise, the peripheral gain parameter in Equation (3) can exceed 1.0. The estimated value of the periphery gain provides an estimate of the proportion of the total variance due to correlated noise:

$$\frac{\sigma_0^2}{\sigma^2 + \sigma_0^2} = 1 - \hat{g}_p^{-2} \qquad (4)$$

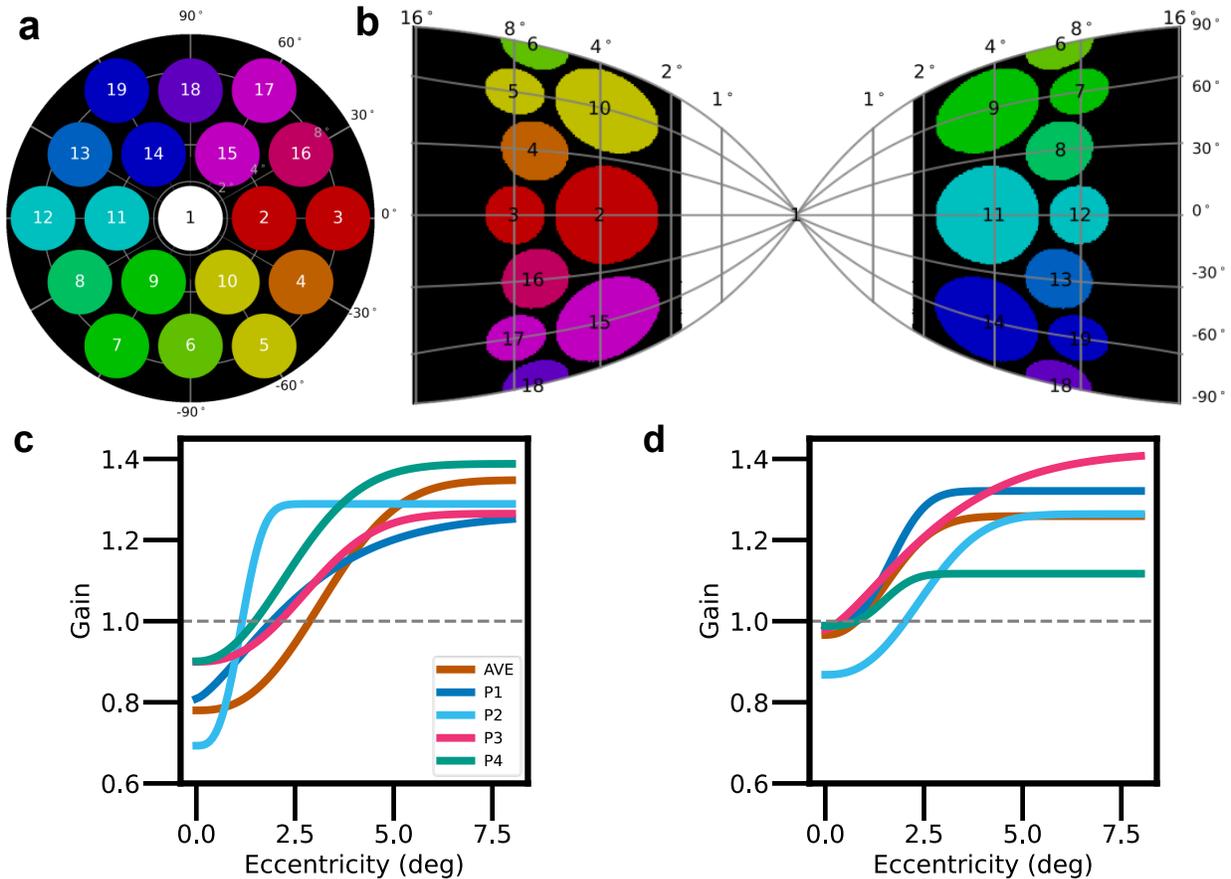

**Figure A3** Attentional gain model parameters for individual observers. **a**. Spatial location of background patches in image space. **b**. Spatial location in V1 space. Colors indicate the direction of a location from the display center. Iso-orientation and iso-eccentricity contours are marked with gray lines. **c**. Estimated attentional gain as a function of retinal eccentricity in the 19-location search task, for the average human observer and the four individual observers, in the same order as the four rows in Figure A1. Estimated foveal and peripheral gain ($g_f, g_p$): AVE (0.780, 1.348), O1 (0.809, 1.262), O2 (0.693, 1.289), O3 (0.900, 1.265), O4 (0.901 1.388). **d**. Estimated attentional gain in the 7-location search task. Estimated foveal and peripheral gain ($g_f, g_p$): AVE (0.966, 1.259), O1 (0.989, 1.321), O2 (0.868, 1.264), O3 (0.995, 1.261), O4 (0.983, 1.240).

## Heuristic detectability maps

A heuristic covert searcher is any model that does not make decisions exactly as the optimal searcher. We focus on a family of models that uses simplified (heuristic) priors and detectabilities in max rule:

$$\hat{\mathbf{x}} = \arg\max_{\mathbf{x}} \left\{ \ln \hat{p}(\mathbf{x}) + \hat{d}'(\mathbf{x}) \left[ R'(\mathbf{x}) - 0.5\hat{d}'(\mathbf{x}) \right] \right\} \quad (5)$$

where $\hat{p}(\mathbf{x})$ and $\hat{d}'(\mathbf{x})$ are the heuristic prior and detectability maps.

The simulated $d'$ maps considered here were of the form

$$d' = \frac{d'_{max} e_2}{e + e_2} \quad (6)$$

Where $d'_{max}$ is the peak value of $d'$ in the center of the fovea, $e$ is eccentricity ($e = \|\mathbf{x}\|$), and $e_2$ is the eccentricity at which $d'$ reaches half the value in the center of the fovea.

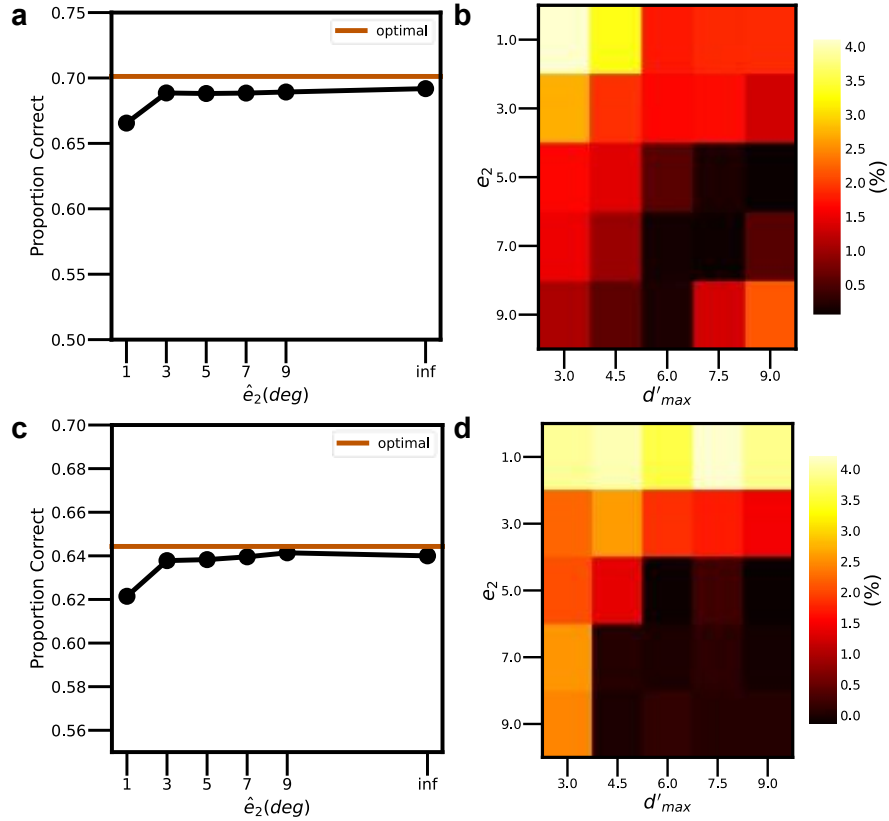

**Figure A4** Comparison of optimal and heuristic searchers. **a**. The overall proportion correct over 25 conditions with $d'_{max}$ = 3, 4.5, 6, 7.5, 9 and $e_2$=1,3,5,7,9, for the optimal searcher (orange), and heuristic searchers (black). The optimal searcher uses the optimal decision rule in each condition. The heuristic searchers use a fixed assumed $d'$ map across all 25 conditions. For each heuristic, the fall-off rate $e_2$ was first fixed, and then the $d'_{max}$ was fitted to maximize overall proportion correct across all 25 conditions. The target-absent prior was 0.5. **b**. The heatmap of the performance lag, defined as the difference between the proportion correct of the optimal searcher and that of a fixed heuristic searcher ( $d'_{max}$ = 6.9, $e_2$ = 7.0). The optimal searcher uses the optimal decision rule in each condition. The single heuristic searcher has the values of $d'_{max}$ and $e_2$ that maximize overall proportion correct across all 25 conditions. **c**. The overall proportion correct plot as in **a**, but with a target-absent prior of 0.0. **d**. The heatmap of the performance lag, as in **b**, but with a target-absent prior of 0.0.

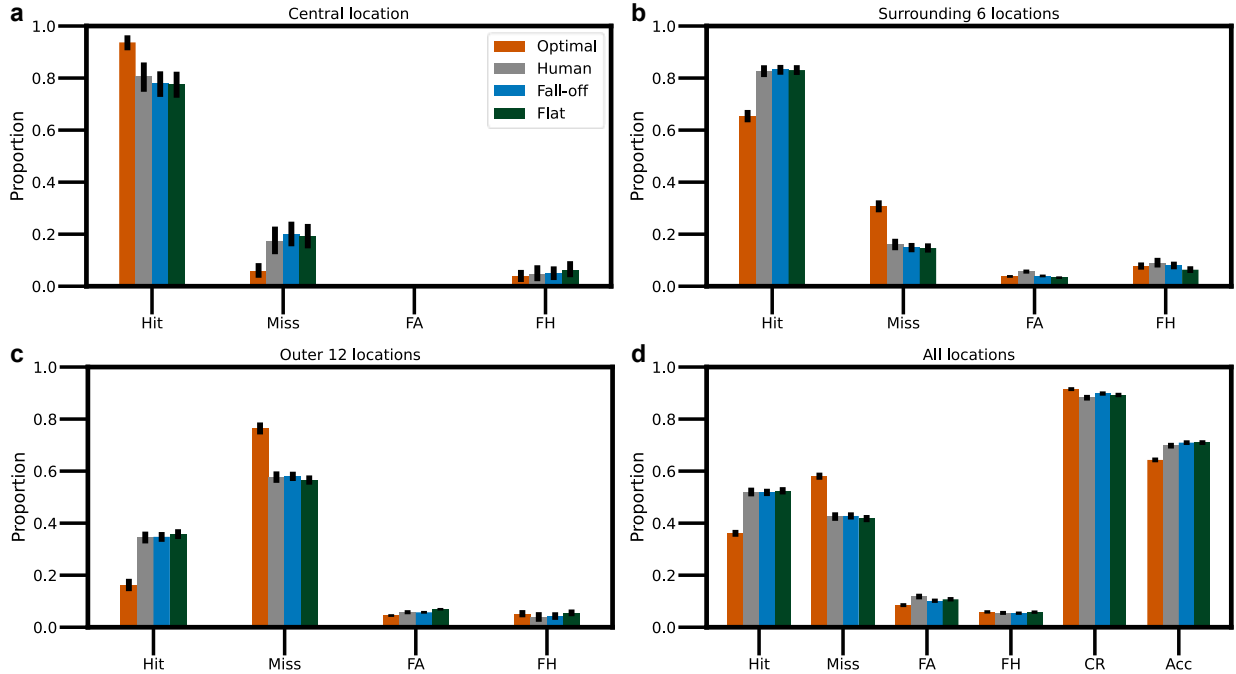

**Figure A5** Correct and incorrect responses in the 19-location search task by retinal eccentricity. **a**. Histograms of hits, misses, false alarms and false hits for the central location, for the average observer (gray), the optimal observer assuming statistical independence of responses across the 19 locations (orange), the optimal heuristic model observer that allows a fall-off in the $d'$ map (blue, $g_f$=0.780, $g_p$=1.348, $a$=3.781, $b$=2.580, $d'_{max}$ = 3.5, $e_2$ = 7.0), and the optimal heuristic model observer with a flat $d'$ map (dark green, $g_f$=0.761, $g_p$=1.342, $a$=3.231, $b$=1.599, $d'$= 3.0). **b**. Histograms for the surrounding six locations. **c**. Histograms for the outer twelve locations. **d**. Histograms for all locations; correct rejections and overall accuracy are also included. (Error bars are bootstrapped 95% confidence intervals.) The fall-off heuristic: log-likelihood = -11758, AIC = 23529. The flat heuristic: log-likelihood = -11787, AIC = 23584. The optimal searcher: log-likelihood = -12039, AIC = 24078. The fall-off model is $e^{27.5}$ times as probable as the flat heuristic model and $e^{274.5}$ times as probable as the optimal searcher.

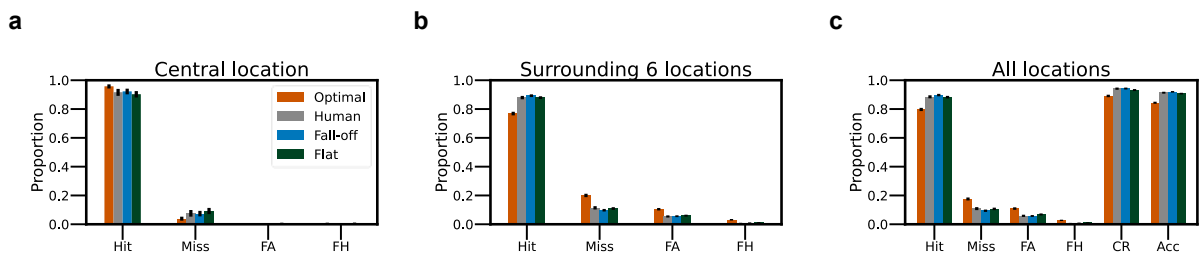

**Figure A6** Correct and incorrect responses in the 7-location search task by retinal eccentricity. **a**. Histograms of hits, misses, false alarms and false hits in the central location, for the average observer (gray), the optimal observer assuming statistical independence of responses across the 7 locations (orange), the optimal heuristic model observer that allows a fall-off in the $d'$ map (blue, $g_f$=0.966, $g_p$=1.259, $d'_{max}$ = 4.6, $e_2$ = 15.8), and the optimal heuristic model observer with only a constant $d'$ map (dark green, $g_f$=0.860, $g_p$=1.325, $d'$ = 3.7). **b**. Histograms for the surrounding six locations. **c**. Histograms for all locations; correct rejections and overall accuracy are also included. (Error bars are bootstrapped 95% confidence intervals.) Comparing the parameter values here with those in Figure S2, we see that the estimated levels of foveal neglect and correlated noise are slightly lower here. The all-off heuristic: log-likelihood = -8228, AIC = 16468. The flat heuristic: log-likelihood = -8230, AIC = 16470. The optimal searcher: log-likelihood = -8405, AIC = 16810. The fall-off model is $e^{1.0}$ times as probable as the flat heuristic model and $e^{171}$ times as probable as the optimal searcher.

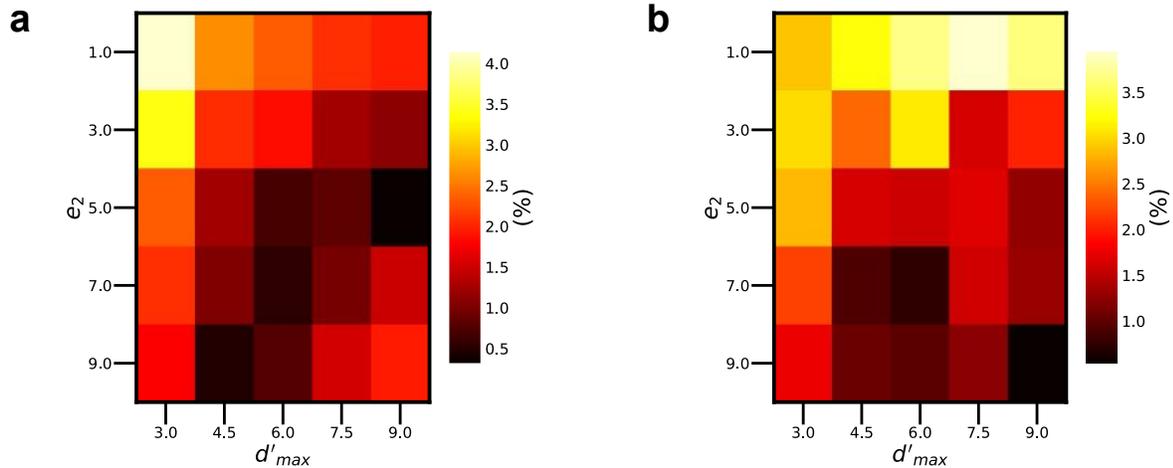

**Figure A7** Comparison of the optimal searcher and a heuristic searcher for random $d'$ maps. **a**. The heatmap of performance lag, which is defined as the difference between the proportion correct of the optimal searcher and that of the heuristic searcher. There are 25 conditions in total, where the baseline $d'$ map is all possible combinations of the following map parameters: $d'_{max}$ = 3, 4.5, 6, 7.5, 9 and $e_2$ =1,3,5,7,9. In each trial, the actual $d'$ map is a sample of the multi-variate independent Gaussian distribution, with the baseline $d'$ map specifying the mean $d'$ at each location and the standard deviation of the $d'$ value being 20% of the mean $d'$ value. The optimal searcher uses the exact random $d'$ map on every trial. The heuristic searcher uses a fixed map with values of $d'_{max}$ and $e_2$ that maximize overall proportion correct across all 25 conditions ($d'_{max}$ = 6.9, $e_2$ = 7.0). The target-absent prior was 0.5. The average lag across all conditions is 1.57%. **b**. The heatmap of the performance lag of the same heuristic searcher for the case where the target-absent prior was 0.0. The average lag across all conditions is 2.08%.

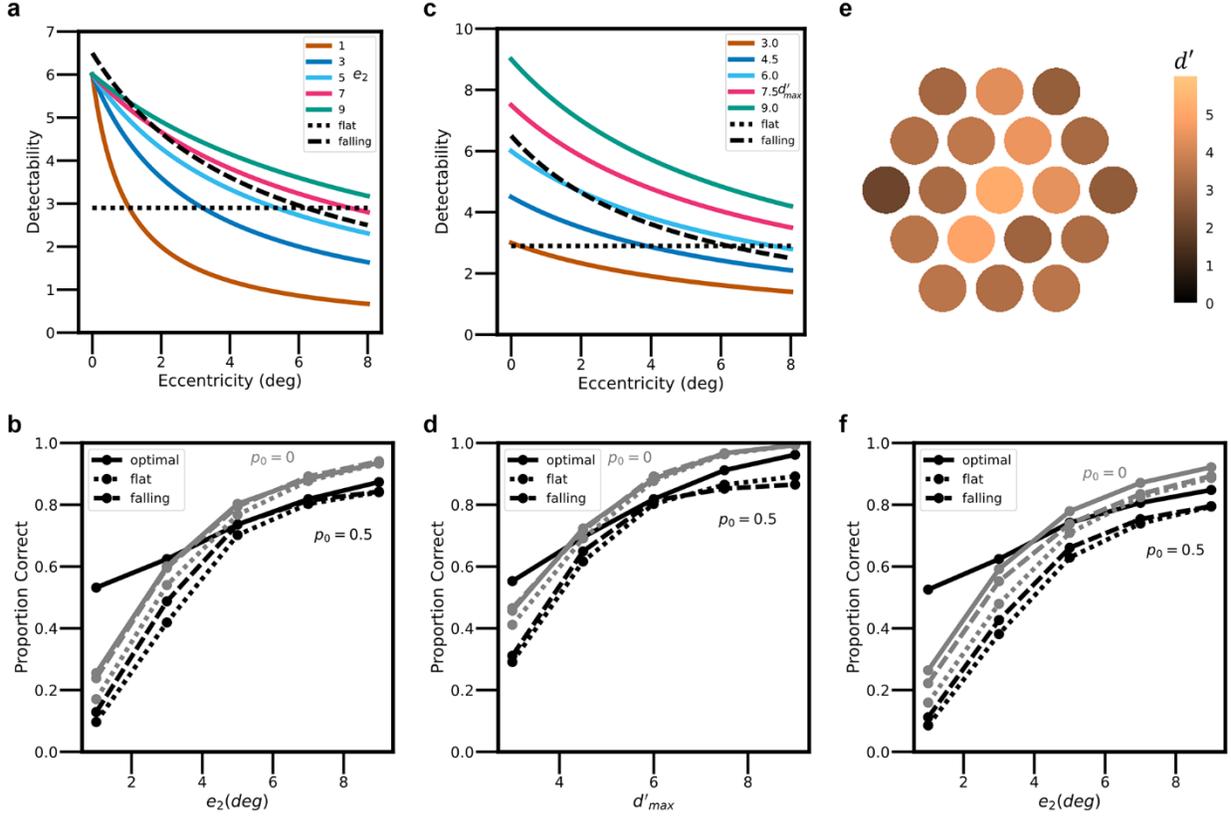

**Figure A8** Heuristic searchers without normalization. In these simulations we assume that the differences in detectability with location are due to differences in the standard deviation (contrast) of the white noise. Specifically, we assume that the mean response when the target is present is always 1.0 and when absent is 0.0, and that the standard deviation at location $k$ is $\sigma_k$. Thus, the detectability is one over the standard deviation: $d'_k = 1/\sigma_k$. When these responses are normalized by the standard deviation, the optimal decision rule is still given by Equation 1 and the heuristic performance is still that shown in Figure 4. If the responses are not normalized, then the optimal decision rule is

$$\hat{k} = \arg\max_{k \in [0,n]} \left( \ln p_k + d'^2_k R_k - 0.5 d'^2_k \right)$$

and the heuristic decision rule is

$$\hat{k} = \arg\max_{k \in [0,n]} \left( \ln p_k + \hat{d}'^2_k R_k - 0.5 \hat{d}'^2_k \right)$$

The solid curves in **b**, **d** and **f** show the overall accuracy for the optimal decision rule; these curves are identical to those in Figure 3. The dotted and dashed curves show the overall accuracy for the same heuristic $d'$ maps as in Figure 4, except the heuristic maps were scaled differently here to maximize overall accuracy. As can be seen, the heuristics perform substantially worse without normalization. Also, note that the means and standard deviations of the responses in these simulations can be scaled by an arbitrary constant without affecting the results.

**Heuristic prior-probability maps**

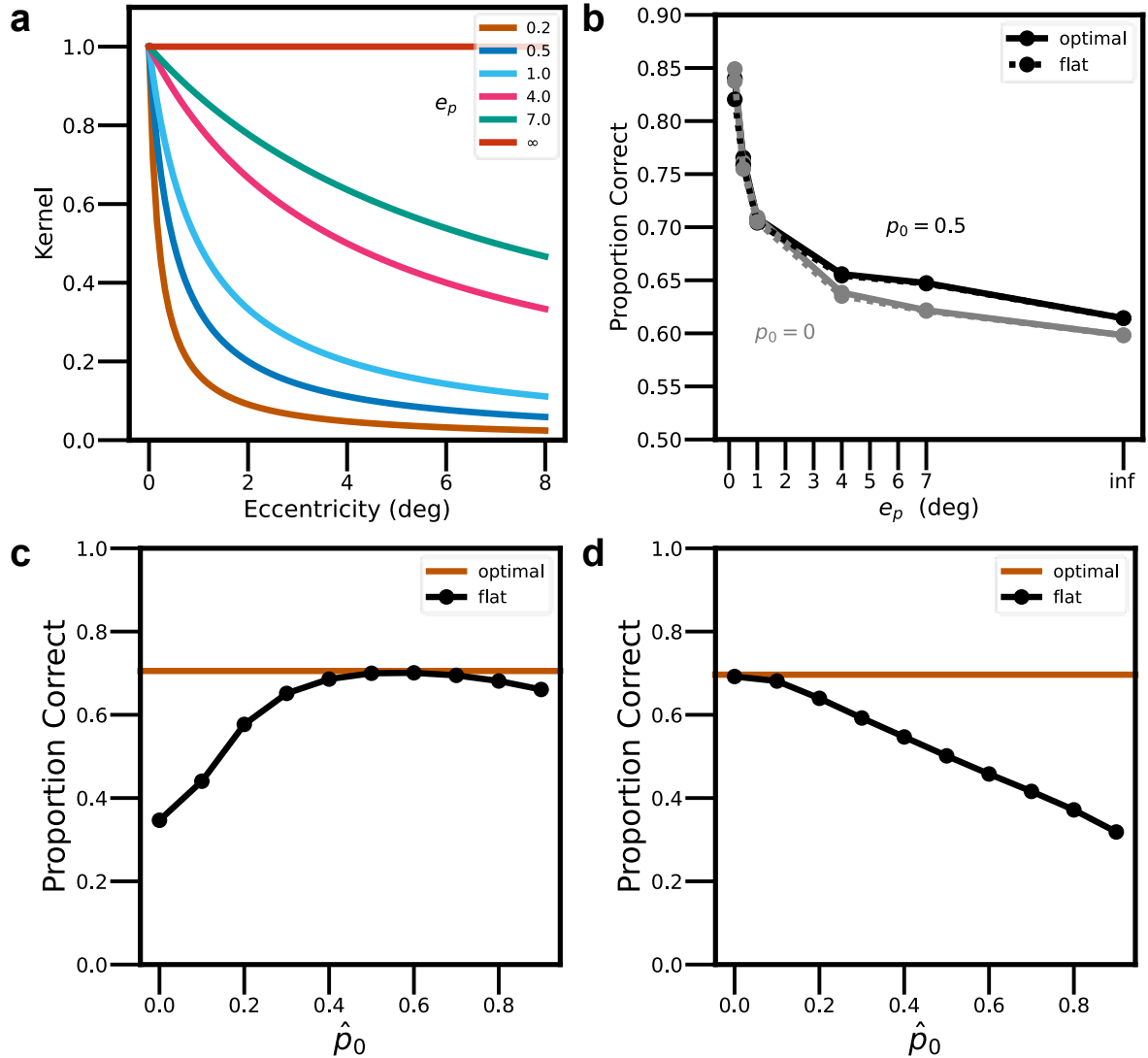

**Figure A9** Effect of heuristic priors on search performance in the task where the observer reports the target location or that the target is absent. **a**. The actual prior probability falloff as a function of retinal eccentricity $e$, for 6 different values of the falloff parameter $e_p$, $p(e) \propto (1-p_0)e_p/(e+e_p)$, where $p_0$ is the prior probability of target absent. **b**. Solid curves are the search accuracy of the optimal searcher as a function of $e_p$, for a target absent priors of 0.5 (black) and 0.0 (gray). The dashed curves are the search accuracy for the heuristic assumption of a flat prior over target location but the correct target absent prior. The actual $d'$ map had a $d'_{max}$ of 5 and a $e_2$ of 4; no heuristics in $d'$ map were assumed. **c**. The average overall accuracy across all 6 conditions of the optimal (orange line) and heuristic searchers (black curve) that assume a flat prior over target location and assume various target-absent prior values $\hat{p}_0$, when $p_0 = 0.5$. Accuracy remains near optimal for $0.4 < \hat{p}_0 < 0.7$. **d**. The average overall accuracy across all 6 conditions of the optimal (orange line) and heuristic searchers (black curve) that assume a flat prior over target location and assume various target-absent prior values $\hat{p}_0$, when $p_0 = 0.0$. Accuracy remains near optimal for $0.0 < \hat{p}_0 < 0.1$.

33